# Prediction of Protein Aggregation Propensity via Data-driven Approaches

Seungpyo Kang[1,+], Minseon Kim[1,+], Jiwon Sun[1,+], Myeonghun Lee[2,*], and Kyoungmin Min[1,*]

[1]School of Mechanical Engineering, [2]School of Systems Biomedical Science, Soongsil University, 369 Sangdo-ro, Dongjak-gu, Seoul 06978, Republic of Korea

## ABSTRACT

Protein aggregation occurs when misfolded or unfolded proteins physically bind together, and can promote the development of various amyloid diseases. This study aimed to construct surrogate models for predicting protein aggregation via data-driven methods using two types of databases. First, an aggregation propensity score database was constructed by calculating the scores for protein structures in Protein Data Bank using Aggrescan3D 2.0. Moreover, feature- and graph-based models for predicting protein aggregation have been developed using this database. The graph-based regression model outperformed the feature-based model, resulting in $R^2$ of 0.95, although it intrinsically required protein structures. Second, for the experimental data, a feature-based model was built using Curated Protein Aggregation Database 2.0, to predict the aggregated intensity curves. In summary, this study suggests the approaches that are more effective in predicting protein aggregation, depending on the type of descriptor and the database.

[+]These authors equally contributed to this work.
[*]Corresponding author: leemh216@gmail.com (M. Lee), kmin.min@ssu.ac.kr (K. Min)

# INTRODUCTION

Protein aggregation occurs when misfolded or unfolded proteins physically bind together due to aging, mutations, or environmental stress.[1–5] Undesirable protein aggregation causes neurodegenerative disorders, such as Alzheimer's disease, Huntington's disease, and Parkinson's disease.[6–9] Moreover, studying protein aggregation is essential to understand the development of protein-based pharmaceuticals against cancer, autoimmune diseases, and metabolic disorders.[10–12] Consequently, understanding protein aggregation is important for promoting human health via preventing neurodegenerative diseases and developing novel therapeutic methods.

Over the past few decades, protein aggregation has been better understood because numerous databases have accumulated data about the protein sequence causing aggregation and clinical data from patients suffering from protein aggregation-related diseases. Representative studies related to this are as follows. (1) The ProADD database contains information on 600 proteins involved in 12 protein aggregation diseases.[13] (2) The AmyLoad website-based database contains 1,300 unique sequences that merge amyloidogenic and non-amyloidogenic sequences from several databases (WALTZ-DB[14], AmylHex[15], and AmylFrag[16][17]). (3) The database of Parkinson's disease (PDbase) comprises information on 2,698 genes related to Parkinson's disease from clinical data.[18] Moreover, protein aggregation has been measured via various methods as follows: (1) TANGO predicts protein aggregation based on physicochemical principles.[19] (2) Aggrescan3D predicts amyloid-forming sequences using a position-specific scoring matrix.[20] (3) AggScore predicts aggregation-prone regions in protein using a 3-dimensional (3D) structure input.[21] Consequently, several databases and algorithms are available to help analyze protein aggregation. However, despite the higher availability of data and methodologies, the quantity of data substantially falls short of the number of identified proteins, and methodologies for predicting protein aggregation can be time-consuming when applied to many newly discovered proteins.

Many recent studies have attempted to predict the aggregation behavior of various proteins quickly and accurately using machine learning (ML). While predicting aggregation through ML, many studies have predicted models depending on the target, such as the aggregation-prone region (APR), which is a specific region of the amino acid sequence for polypeptide chain prediction and aggregation propensity (AP) score prediction.[22] To predict the APR and AP, sequences can be inputted to the ML and deep learning (DL) models. Example cases are: (1) ANuPP (a versatile tool to predict the aggregation nucleating region in peptides and proteins) predicts the APR using atomic-level characteristic features using an ensemble classifier, which shows 77% accuracy on 10-fold cross-validation.[23] (2) Prediction of the aggregation rate and classification between enhancers or mitigators using a sequence feature-based method (dataset of 220 unique mutations in 25 proteins from experiments) and SVM, which showed an overall accuracy of 69%.[24] (3) RFAmyloid (a web server for predicting amyloid proteins) predicts amyloid proteins using random forest, which shows 89.19% accuracy and two feature extraction methods (188-Dimensional feature extraction method and pse-in-one feature extraction method).[25] However, in the case of this sequence feature-based model, the protein 3D structural information has some drawbacks; for example, the protein folding and native state cannot be considered. Therefore, a novel model must be developed that considers automated feature extraction from protein sequences and structural information.

Unfortunately, the lack of experimental data and an established accurate methodology remains unresolved. However, if a sufficient and high-quality database exists for protein aggregation, it can be applied to powerful ML and DL. Therefore, in this study, to secure sufficient data, Aggrescan3D 2.0 (A3D) was used to build a static and dynamic database of all 144,612 protein structural data that could be calculated in the Protein Data Bank (PDB), as is shown in **Figure 1**.[26, 27] Moreover, sequence feature- and structural graph-based models for predicting AP scores were constructed using the established database. Furthermore, to effectively predict protein aggregation using the accumulated experimental database, surrogate models predicting aggregation intensity using Curated Protein Aggregation Database 2.0 (CPAD) were constructed.[28] Overall, this study proposes a comparison of both computational and experimental approaches for constructing a database for effectively predicting protein aggregation, particularly feature-based and graph-based methods for computational approaches.

# METHODS

## 1. Computational approach

To construct an initial computational database for protein aggregation, the AP score was calculated using previously developed A3D software for PDB structures.[26, 27] As is shown in **Figure 2**, the overall process of the computational approach of this study is as follows. (1) The AP and structural stability of the PDB structures were analyzed using A3D calculations in static and dynamic modes. Protein flexibility was further reflected in the A3D dynamic mode, as is shown in **Figure 2(a)**. (2) The feature- and graph-based methods are applied to the constructed database. **Figure 2(b) and 2(c)** show the processes of the feature-and graph-based models, respectively. Therefore, the propensity for protein aggregation was predicted and the prediction accuracy was compared using these two approaches.

### 1.1 Database construction

In the computational approach, the previously developed A3D was used to determine the protein aggregation propensity for each PDB structure.[26, 27] A3D calculates the aggregation propensity of a protein based on its amino acid sequence and 3D structure. In the A3D package, static-mode calculations consider physicochemical properties to calculate the aggregation propensity in the static conformation. It is useful for predicting the aggregation propensity of proteins that are structurally characterized using experimental techniques such as X-ray crystallography and nuclear magnetic resonance (NMR) spectroscopy. Dynamic-mode calculations utilized the FoldX force field to minimize the structural configuration of the protein.[29] The aggregation propensity was then calculated up to 12 times for each potential point mutant. This process considers the dynamic characteristics of proteins at minimum cost. In summary, the static mode is capable of fast calculation analysis of the static conformation, whereas the dynamic mode is slow, but can include various dynamic characteristics. To utilize all the advantages and prediction performance variations of each calculation method, static and dynamic databases were constructed.

**1.2 Feature-based Model**

Biopython[30], PyBioMed[31], and ProtScale[32] features were generated to sufficiently describe protein aggregation using only a given sequence. Biopython (19) features provide simple protein properties such as molecular weight and aromaticity (**Table S1**). PyBioMed (9,890) features represent the structural and physicochemical features of proteins and peptides derived from amino acid sequences (**Table S2**). These structural and physicochemical features can be used to predict protein structural and functional classes, subcellular locations, peptides with specific properties, and post-translational modifications. ProtScale (76) considers calculated values based on various chemical and physical properties of each amino acid. The values for each amino acid in ProtScale and detailed descriptions are provided in **Supporting Information (SI)** as CSV file. Biopython features were used for simple protein analysis, PyBioMed features were used to consider various structural and physicochemical properties, and ProtScale features were used to consider the values of each amino acid calculated based on various properties. Consequently, for the computational approach, 1,435 features were used to construct a feature-based model: Biopython (19), PyBioMed (1,340), and ProtScale (76). To reduce computational costs and solve generation errors, 8,550 PyBioMed features were excluded from the feature generation process, including Tripeptide Composition (8,000), Moran Autocorrelation (240), Geary Autocorrelation (240), PAAC (30), and APAAC (40). Because the tripeptide composition can be considered with more detailed information through the Dipeptide Composition, it was excluded to reduce computational costs. Moreover, the Moran Autocorrelation, Geary Autocorrelation, PAAC, and APAAC were excluded to obtain additional data without generating errors.

Regression and classification models are constructed to consider various criteria that can be applied to static and dynamic databases. For model comparison, five-fold cross-validation was performed. Details of the model comparison are as follows: (1) eight regression models; linear regression (LR), ridge (RIDGE), lasso (LASSO), random forest (RF), gradient boosting (GB), light gradient boosting machine (LGBM), and extreme gradient boosting (XGB) regressors; (2) six classification models; Naïve Bayes (NB), LR, RF, GB, and XGB classifiers. After model selection, to construct a surrogate model, hyperparameter optimization was performed with only 25% random sampling using a grid-search algorithm to reduce excessive computational cost.

**1.3 Graph-based Model**

To apply the latest deep learning method used in many protein research fields, a graph convolution neural network (GCN) was adopted using PyTorch Geometric packages,[33] and for the graph convolution algorithm, GCNConv was implemented.[34] As is shown in **Figure 3(a)**, a graph representation of the protein was generated from the 3D atomic coordinates of the PDB files. The graph representation (G) can be marked as G = (V, E), where V denotes the nodes (each amino acid), and E denotes the edges (indicating whether a connection exists between each amino acid). In the present study, E indicated the presence of a connection between two amino acids if the alpha carbon level was less than the threshold ($7\dot{A}$). Protein graph G is encoded as a node feature matrix X and an adjacency matrix A. The shape of X is a 20-dimensional node feature matrix (N × 20, where N is the number of amino acids in the protein) because the one-hot encoding of 20 types of amino acids was used to represent X. On the other hand, the shape of A is a square (N × N) with the length of one side being the length of N because A was generated by expressing the connection relationship of each amino acid (connected as 1 and disconnected as 0).[35]

As is shown in **Figure 3(b)**, the GCN model is mainly divided into a graph convolution layer, which is the step for extracting the general features from the protein graph, and a dense layer, which is the step for predicting the A3D score values or classes based on the average of the A3D scores. In this study, both the graph convolution and dense layers are composed of five layers that have synchronized dimensions with the input dimensions for easily adjusting the dimension. The number of input dimensions was set to eight. The configurations of the convolution and prediction layers in the neural networks were similar in both the classification and regression models. However, the output shape, loss of function, and final activation function differed between the classification and regression models. In the classification model, the loss function is a binary cross-entropy (BCE) loss because the shape of the output is a 1-dimension probablity, and the final activation function is a sigmoid function. However, in the regression model, the loss function is the mean squared error (MSE) loss because the shape of the output is a 1-dimension scalar, and the final activation function is a linear function representing the negative value of the A3D score. In the convolution layer, batch normalization (BN) was adopted to avoid the internal covariant shift problem.[36] The BN layer was located before the rectified linear unit (ReLU) activation function. In the dense layer, the ReLU activation functions were adapted, and a dropout layer with a rate of 0.2 was adapted in each layer to prevent overfitting.[37] Adaptive moment estimation (Adam) was adopted as the optimization function for each model.[38] Hyperparameter optimization was performed with 25% randomly sampled data using the grid-search method to reduce computational cost.

## 2. Experimental approach

### 2.1 Database construction

Although the calculated databases provide theoretical predictions of protein aggregation, their accuracy could vary depending on the underlying computational models. In contrast, experimental databases contain information about proteins that have been shown to aggregate experimentally, and can provide a more accurate representation and value of protein aggregation. By analyzing these data using machine-learning algorithms, researchers can identify patterns and features associated with protein aggregation and make more accurate predictions regarding the aggregation propensity of proteins. CPAD is an experimental open-access database that contains extensive information on the kinetic (aggregation experiments to measure kinetics), structural (structures relevant to protein aggregation), and mechanistic (aggregation-prone regions in proteins) properties of protein aggregation. Among the four datasets from CPAD, this study used the Aggregation Kinetics of Proteins, and the details are given in the part S.2 in **SI**.

The Aggregation Kinetics of Proteins is a database of aggregation kinetic data from various experiments, including 83,098 data points.[28] The intensity, which monitors changes in the concentration of aggregated proteins over time and quantifies the degree of aggregation, was considered as the degree of protein aggregation.[39] Two protein aggregation measures provide insight into the kinetics and thermodynamics of protein aggregation: (1) fluorescence intensity and (2) light scattering intensity. In general, as protein aggregation increases, the fluorescence intensity decreases and the light scattering intensity increases.[40, 41] These results provide insight into the kinetics and thermodynamics of protein aggregation. **Figure 4** shows the overall process of aggregated intensity prediction using an experimental database. The aggregated intensity was predicted using polynomial and multi-output regression models as feature-based methods. A corresponding intensity existed for each data point in each experiment, and the intensity values held in each experiment were averaged to facilitate intensity comparison according to the experiment. As is shown in **Figure S1(a)**, the average intensity distribution in each experiment was severely skewed, necessitating the removal of outliers. The average intensity value for each data point is shown in **Figure S1(b)**. For the average intensity, MIN was 0, MEAN was 406, and MAX was 46,396. STD was confirmed to be 1,691, with a large deviation. As the intensity used different arbitrary units for each experiment, no unit was obtained, and unification was impossible. The sequence length provided by the CPAD and the sequence length corresponding to the UniProt ID differed. In the actual experiments, some of the total sequence length was used, and some of the amino acids in the sequence were changed. Because the amount of experimental data was too extensive to apply the sequence modifications of all experimental papers, only experiments with the same UniProt ID sequence length were used for machine learning. Unbalanced data distribution hinders machine learning prediction; therefore, outliers were removed using the interquartile range (IQR), a widely used easy-to-interrupt outlier-removing method.[42]

**2.2 Polynomial regression**

A polynomial regression algorithm was used to unify the data points of intensity for each experiment and to use it as the target value of the multi-output regression model. The final goal was to construct a multi-output regression model that predicts the intensity of the 10 points by dividing the time range of each experiment into 10 equal intervals. As is shown in the machine learning part of **Figure 4**, polynomial regression is more useful for curved data than simple linear regression and estimates the best approximation of the relationship between the dependent and independent variables.[43] Considering that most intensity data over time graphs are curved, we used polynomial curve fitting to predict 10 points of intensity. Ten data points predicted by the polynomial curve-fitting model were used as target values for the multioutput regression model. The multitarget approach has better computational efficiency compared to the single-target approach and ensures better predictive performance when the targets are correlated.[44] The multi-output regression model used five-fold cross-validation as a feature-based prediction model. All multi-output performances are the average validation-set performance values of the five models.

**2.3 Feature-based model**

A feature-based model was used as the input for the multi-output regression model, and the sequence was extracted based on the UniProt ID. The aggregation kinetics of the protein database provided both sequence length and UniProt ID; however, the sequence was modified for each experiment in some cases. Therefore, the sequence length of the database and the sequence length according to the UniProt ID of the database were different. Therefore, for 947 experiments with inconsistent sequence lengths, we decided to use only 132 experimental results with matching sequence lengths because the change in positions of the amino acid groups were uncertain. All possible 9,984 features were obtained using 132 sequences of experimental data from Biopython (18 features), PyBioMed (9,890 features), and ProtScale (76 features). In the A3D section, features that cannot be computed or require a long calculation time were avoided; however, in this section, features that can be constructed were secured as much as possible to identify characteristics that substantially affect model prediction. Among the experimental conditions included in the Aggregation Kinetics of Proteins database, five features, including ndatapoints, temperature, pH concentration, mutation type (including wild, single, double, multiple, triple, and modified type), and time unit (including seconds, minutes, hours, and days) were added to improve predictive performance, and 9,989 input features were used.

# RESULTS AND DISCUSSION

## 1. Computational approach

### 1.1 Database construction

As was previously discussed, an initial computational database was constructed by calculating the AP scores for 144,612 PDB structures using the A3D package. Static and dynamic databases containing the mean AP scores of the amino acids in each structure were constructed. The distribution of the static database is shown in **Figure 5(a)**. Considering the various criteria that can be applied to static and dynamic databases, the class was divided based on the average value (static database: -0.746, dynamic database: -0.206): aggregate (1) and non-aggregate (0), as shown in **Figure 5(b)**. The distribution and class of the dynamic database are obtained in the same manner, as is shown in **Figure 5(c), (d)**.

### 1.2 Feature-based model selection

For both regression and classification models, we considered the models that were more sensitive and suitable for the generated features. For this purpose, five-fold cross-validation was performed on the entire database, and the optimal model was selected. **Figure S2** compares the regression models applied to both static and dynamic databases. Considering $R^2$ and computational efficiency, the XGB model was selected for training both the static and dynamic databases. XGB model performed with $R^2$ of 0.84 (±0.005) and 0.90 (±0.005) for the static and dynamic datebase, respectively. Comparison of classification models proceeded by considering F1-score and conformity with high-dimensional data, and the XGB model was selected, as is shown in **Figure S3**. For each static and dynamic database, XGB model perfomed with an F1-score of 0.88 (±0.006) and 0.90 (±0.005), respectively. Consequently, to consider the performance and suitability of high-dimensional data, the XGB model was selected as the optimal model for both regression and classification. After model selection, hyperparameter optimization was performed on the selected model using 25% of the randomly sampled data to reduce the computational costs, which is shown in **Figure S4**. We note that the densities of 25% of the randomly sampled data and the entire database are the same, which shows that the optimized hyperparameter can be applied reliably enough to the entire database. The hyperparameters used in the regression and classification models are shown in **Table S3 and S4**, respectively.

**1.3 Graph-based model selection**

The hyperparameter optimization process of the GCN model was performed in a manner similar to that of the feature-based model using 25% randomly sampled data. In the GCN model, the training and test losses of each model were also considered to avoid overfitting. **Table S3 and S4** list the best parameters and means of the metrics of the regression and classification models, respectively. During hyperparameter tuning of the GCN model, the input dimensions were 8, 16, and 32. Only graph dimensions of 8, 16, and 32 were considered in this study because previous research has shown that increasing the dimensions of the graph does not always result in optimal performance and prevents overfitting.[45] Consequently, during hyperparameter optimization, the classification model easily becomes overfitted when the dimensions increase. However, in the regression model for both the static and dynamic databases, an input dimension of 32 yielded the best performance. Therefore, the input dimensions of the GCN model were selected as 32 and eight for the regression and classification models, respectively. During hyperparameter optimization of regression models, the best-parameter models performed $R^2$ with 0.82 (±0.002), 0.94 (±0.002) in static and dynamic databases, respectively. For classification, the best-parameter model showed F1-score of 0.85 (±0.001) and 0.91 (±0.004) from the static and dynamic databases, respectively. The learning process for each loss function is shown in **Figure S5**.

**1.4 Comparison of feature-based and graph-based models**

Constructing feature-based models is more straightforward because only sequence information is required for feature engineering. This is advantageous in terms of time and cost because X-ray biocrystallography is not required for representing the protein using its structure. However, structural information such as protein folding and native state cannot be considered. Such drawbacks can be resolved using the structural descriptors of known protein structures for the feature-based model; however, this will complicate the model training process, which was not considered in this study. On the other hand, graph-based models are more sophisticated to build, but they offset the shortcomings of feature-based models by reflecting more comprehensive structural information. Fortunately, with the recent development of AlphaFold[46], a tool capable of accurately predicting protein structures from sequence information, the potential for utilizing graph-based models is expected to intensify.

The prediction results of the surrogate regression models for each static and dynamic database are shown in **Figure 6(a), (b)**. The $R^2$, root mean square error (RMSE), and mean absolute error (MAE) of each surrogate model are shown in **Figure 6(c)**. The feature- based and graph-based models for static database performed with $R^2$ of 0.88 and 0.87, respectively. For dynamic databases, feature-based and graph-based performed with $R^2$ of 0.93 and 0.95, respectively. Comparing the feature- and graph-based models, the graph-based model performs slightly better. However, as was previously mentioned, the graph-based model requires structural information; therefore, an appropriate model should be considered for the given resource. From the perspective of model interpretation, a feature-based model, which has explanatory power for important features, is advantageous over graph-based models. Therefore, an appropriate model selection is required, depending on the given resource and direction to be analyzed. The regression model using the dynamic database performs slightly better. This was because the AP score obtained through structural relaxation was closer to the AP score of the actual structure. Thus, constructing a database and prediction model based on a stable structure is important.

As was mentioned previously, feature-based models can be used for model interpretation. The features that have a profound effect on model training. Using a feature-based model and the previously developed Shapley additive explanation (SHAP) analysis[47], features that were significantly applied to predict protein aggregation were identified (**Figure 7**). As a result of the SHAP analysis for the regression model, the most influential features for each static and dynamic database were as follows: (1) static database, ProtScale|Hydrophilicity, PyBioMed|CTD_SolcentAccessibility, and ProtScale|Rentention_coefficient_HPLC_74; (2) dynamic database, Biopython|Gravy, ProtScale|Hydrophobicity_indices_HPLC_75, and

PyBioMed|AAComp_D. The top 20 features of the static database consisted of four Biopython, six PyBioMed, and 10 ProtScale features, whereas those of the dynamic database consisted of five Biopython, 12 PyBioMed, and two ProtScale features. Because the characteristics of each database are different, the top 20 features are slightly different; however, the Biopython features have the highest conversion ratio in both the static and dynamic databases (4/18 and 5/18, respectively). Additionally, gravy-, flexibility-, and hydrophobicity-related features were highly affected by regression models for both static and dynamic databases. Previously, the gravy, flexibility, and hydrophobicity were found to be related to protein aggregation, and gravy-, flexibility-, and hydrophobicity-related features are crucial for predicting the mean AP score.[48–50] In addition, length was found to predict the mean AP score.

Classification models were constructed considering various criteria that could be applied to static and dynamic databases. The confusion matrices of the classification model for each static and dynamic database are shown in **Figure 8(a) and (b)**. The detailed performance of each surrogate model is shown in **Figure 8(c)**. The surrogate models for the static database performed with 0.90 and 0.90 F1-score, respectively. For the dynamic database, the surrogate models were performed with 0.92 and 0.93 F1-score of respectively. Graph-based regression models performed slightly better. As was mentioned previously, this represents the importance of a stable structure for constructing a database and prediction model. SHAP analysis of the classification models is shown in **Figure S6**. Although a slight difference was observed from the regression SHAP analysis due to class, gravy, flexibility-related, and hydrophobicity-related features, they also meaningfully affected the prediction of the mean AP score.

## 2. Experimental approach

The proposed approach outlines a methodology to predict protein aggregation using experimental data. Detailed information regarding the model selection process and comprehensive analysis is shown in S.5 part of the **SI**. Furthermore, as the data were amalgamated from diverse experiments, the unification of units was impractical because of their arbitrary nature. Hence, the accuracy of a prediction model serves as the basis for determining its reliability. To select the optimal degree coefficient of the polynomial model, 1,540 raw datasets were used, as is shown in **Figure S7** to compare $R^2$ and MAE of the polynomial model according to the number of degrees. The 5° polynomial model exhibited the best performance, with $R^2$ = 0.957. **Table S5** presents a comparison of the performance of the data preprocessing methods for improving the performance of the polynomial regression model. IQR+20% model using 1,079 data points exhibited the best performance, with $R^2$ = 0.971. Furthermore, **Figure S8** shows a balanced distribution, with many high-intensity values removed from the IQR+20%. By comparing the data distribution and the predicted value distribution of IQR+10% and IQR+20% data, high-intensity data were removed when IQR+20% data were used, showing a relatively balanced predicted value distribution without focusing on high values. In conclusion, 10 intensity points from 1,079 experiments were predicted using the IQR+20% dataset via a polynomial model.

To generate features, no data on the entire sequence in the database was present, so the sequence had to be obtained with a UniProt ID. Data with a sequence length equal to that of the UniProt sequence (132 data points) were selected from IQR+20% (1,079 data). The performance of three multioutput regression models, the base model that uses the extra tree regressor (ET) multioutput model, the ensemble model using ensembled with AdaBoosting and ET estimator, and the chain model using ET estimator, was compared to determine which multioutput algorithms have the best performance in **Table S6**. The ET regressor used as an estimator and the AdaBoosting regressor used as a multi-output model were selected using auto ML PyCaret to efficiently compare various models.[51] We attempted to ensemble ET with $R^2$ = 0.715 and AdaBoosting with $R^2$ = 0.617, which ranked high among the comparative models using PyCaret. Chain regression models use a sequential approach to predict target variables. Chain regression models can capture the dependencies between the output variables; however, they may not perform well if the dependencies are complex or nonlinear. To investigate whether a chain model would be more effective, based on the relationship between the ten intensity points, they were incorporated into the comparison target. The chain model had the highest $R^2$ (0.503) for the validation set and the lowest $R^2$ (0.962) for the training set. The base model came second, with a validation-set performance $R^2$ of 0.498, and the training-set performance $R^2$ was 0.973. Despite a relatively minor performance difference between the two models and the higher training performance demonstrated by the base model, the multi-output model was ultimately selected because of its ability to predict all output variables simultaneously using the input variables, thereby exhibiting a faster speed than the chain model that predicts each output variable sequentially. Consequently, a multi-output model employing the estimator, which demonstrates superior speed, was selected.

Feature selection was performed by considering the previously mentioned SHAP value because adding more features could potentially fall into the curse of dimensionality.[52] Feature selection was performed using SHAP analysis, which obtains feature importance based on three properties: local accuracy, missingness, and consistency. **Table S7(a)** shows a list of all the features used, and **Table S7(b)** shows the top five features with the highest feature importance (SHAP value). In **Table S7(c)**, removing low-importance feature models using 500 feature-importance top 5% features showed the highest performance, with an $R^2$ of 0.498 for the validation set. We constructed an efficient prediction model that reduced the number of features from 9, 989 to 500. The hyperparameters obtained from the Bayesian optimization[53] using all features are listed in **Table S8**. Upon tuning the hyperparameters, as is shown in **Table S8(a)**, Bayesian Optimization was performed using all the features. Despite using Bayesian optimization, accuracy did not improve; hence, the default parameter was used. **Table S8(b)** shows the performance of the model using the default parameters.

Methods such as eliminating outliers, experimental feature selection, and hyperparameter tuning were used to apply machine learning to the experimental data; however, the predictive performance of the multi-output model was low. The main reason for this is that unit unification of intensity is not possible, and only 132 very small datasets that match UniProt ID were used because full-text sequences were not provided in the database. If the protein sequence information in the database is unclear owing to length differences or mutations, despite the presence of experimental data, it is difficult to use. In environments with limited resources and very small amounts of data, researchers have sought to demonstrate the optimal choices that can be made when given the experimental data. To illustrate this, four materials are selected as examples. We selected the best performing model among the five multi-output models with five-fold cross-validation and analyzed the top four materials with the best prediction. **Figure 9(a) and (b)** show the results of the polynomial and multi-output regression models, respectively. **Table S9** provides the details of the experimental methods and conditions for the top four proteins. For reference, the detailed descriptions of each protein are as follows: Top 1 protein (T-0129) and top 2 (T-0122) exhibit s-shaped behavior and are alpha-synucleins involved in neurodegenerative maladies, including Parkinson's disease (PD), multiple system atrophy, and dementia with Lewy bodies.[54] The top three proteins (T-1103) exhibited a linear behavior and were islet amyloid polypeptides that cause diabetes mellitus (type II diabetes) when aggregated.[55] The top four proteins (T-0664) exhibited linear behavior and were Galectin-7, its aggregation is involved in the pathogenesis of primary localized cutaneous amyloidosis (PLCA).[56] **Figure 9(c)** shows the validation performances of the polynomial and multi-output regression models for each material.

**3. Comparison of computational and experimental approaches**

This study employed two approaches for developing prediction models depending on the database and surrogate model accessibility: (1) using computational database construction and (2) using an experimental database. Although constructing a computational database can be time-consuming and resource-intensive, it also allows for developing predictive models that can provide faster and simpler predictions than conducting repetitive calculations on a large amount of new data. Moreover, computational databases can help overcome the limitations of experimental data, which require significant costs and effort. In contrast, although experimental databases consist of actual prediction targets and may have higher data quality, the amount of data remaining after quality curation maybe insufficient for effective predictive model development. In this study, the prediction models were verified using both computational and experimental databases, and the two approaches presented different directions for developing protein aggregation prediction models.

**CONCLUSIONS**

In the present study, surrogate models were constructed to predict protein aggregation using data-driven methods. For the initial computational database, the AP score was calculated using the previously developed A3D for 144,612 PDB structures. Consequently, static and dynamic databases with mean AP scores were constructed. To confirm which features were most effective in predicting protein aggregation, feature- and graph-based models were compared. The graph-based regression model outperformed the feature-based model, resulting in $R^2$ of 0.95, although it required protein structures. In addition, we introduced an experiment-based approach to predict aggregation intensity, providing insight into the kinetics, concentration, and thermodynamics of protein aggregation. By applying polynomial and multi-output regressions, we propose a method for effectively predicting the aggregated intensity when various experimental data are available. The surrogate model could be improved by expanding the organized database for protein aggregation. We believe that the following computational and experimental approaches will effectively accelerate protein aggregation by applying them to new structures, such as the AlphaFold Protein Structure Database.

## ACKNOWLEDGMENTS

This work was supported by National Research Foundation of Korea (NRF) grants funded by the Korean government (MSIT) (Nos. 2020R1F1A1066519 and 2022R1F1A1074339). This study was supported by Standigm.

## DECLARATIONS

**Availability of data and materials**

The primary source code used in this study are available at Github:

https://github.com/alstjs0528/Aggregation. The database built in this study is available at Figshare: https://doi.org/10.6084/m9.figshare.22492606.v1.

**Competing interests**

We declare no conflict of interest.

**Funding**

This work was supported by the National Research Foundation of Korea (NRF) grant funded by the Korea government (MSIT) (No. 2020R1F1A1066519, No. 2022R1F1A1074339). This work was supported by Standigm.

**Authors' contributions**

SK and JS contributed data curation, validation, and writing - original draft in relation to the computational approach section. MK contributed data curation, validation, and writing - original draft in relation to the experimental approach section. ML contributed methodology, data curation, writing - review & editing for the entire contents. KM contributed conceptualization, methodology, writing - review & editing, and supervision for this study. All authors reviewed the manuscript.

**Acknowledgments**

The authors would like to thank Seung-Woo Seo for helpful advice on protein aggregation.

**Corresponding author**

Correspondence to Myeonghun Lee and Kyoungmin Min.

# APPENDIX A. SUPPORTING INFORMATION

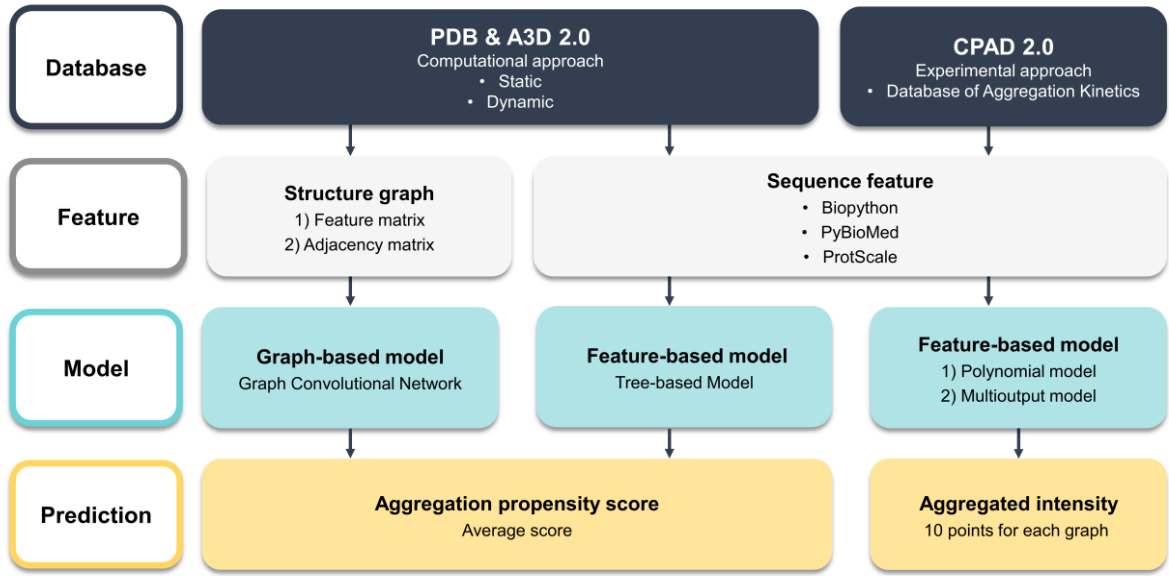

**Figure 1.** Schematic of this study.

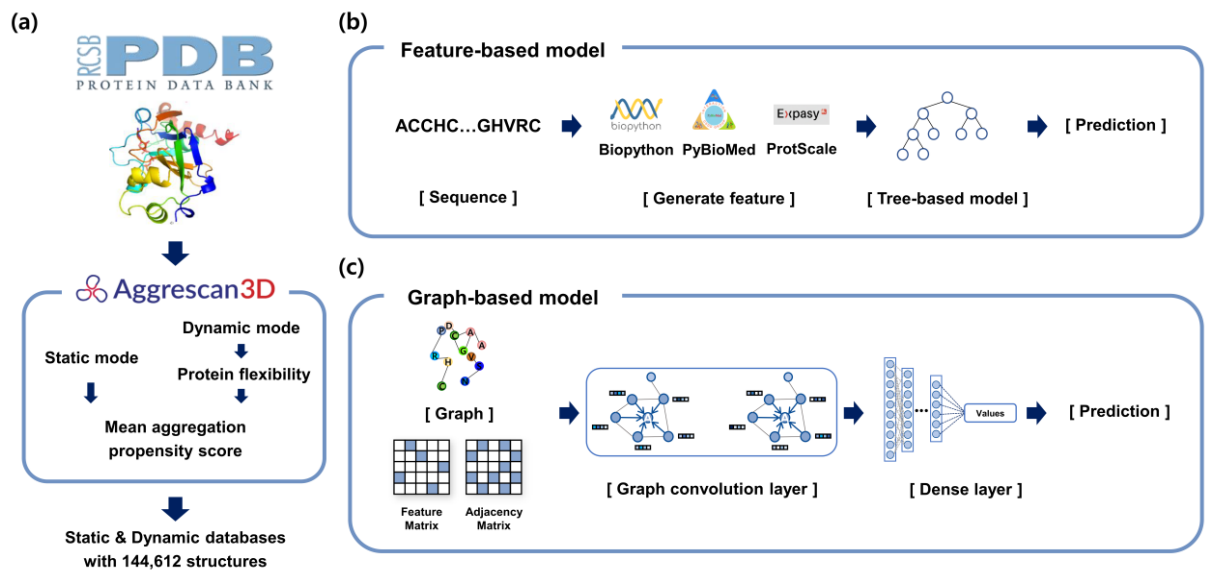

**Figure 2**. Overall process of the computational approach of this study. **(a)** Overall procedure of database construction from A3D calculations with PDB structures. Both sequence-based and graph-based models were constructed to predict mean AP score; **(b)** sequence-based model, **(c)** graph-based model.

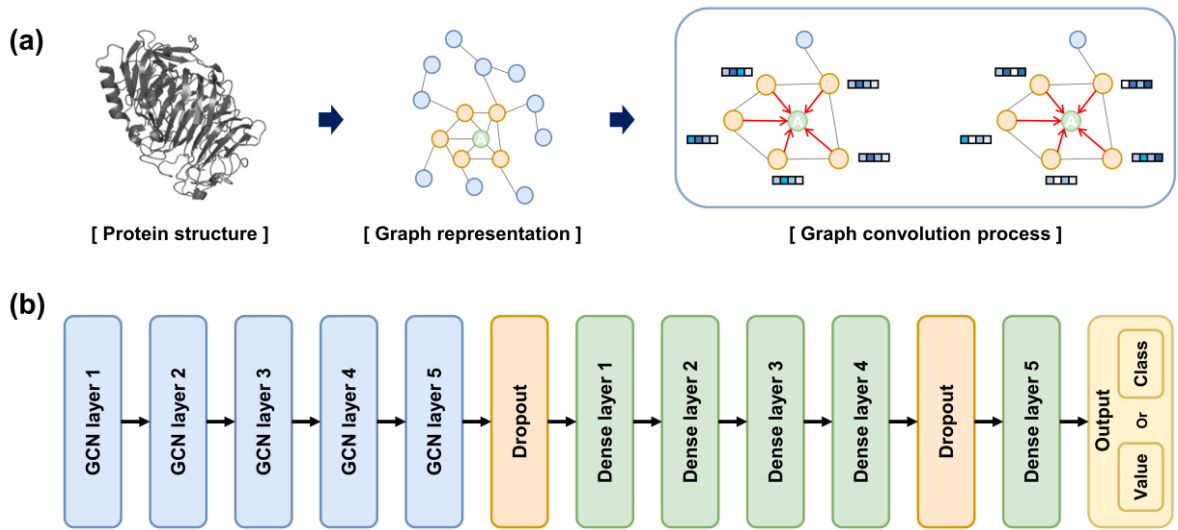

**Figure 3. (a)** Schematic of graph convolution process of GCN model. **(b)** Overview of GCN model architecture. The output is designed differently depending on whether it is a classification model or a regression model.

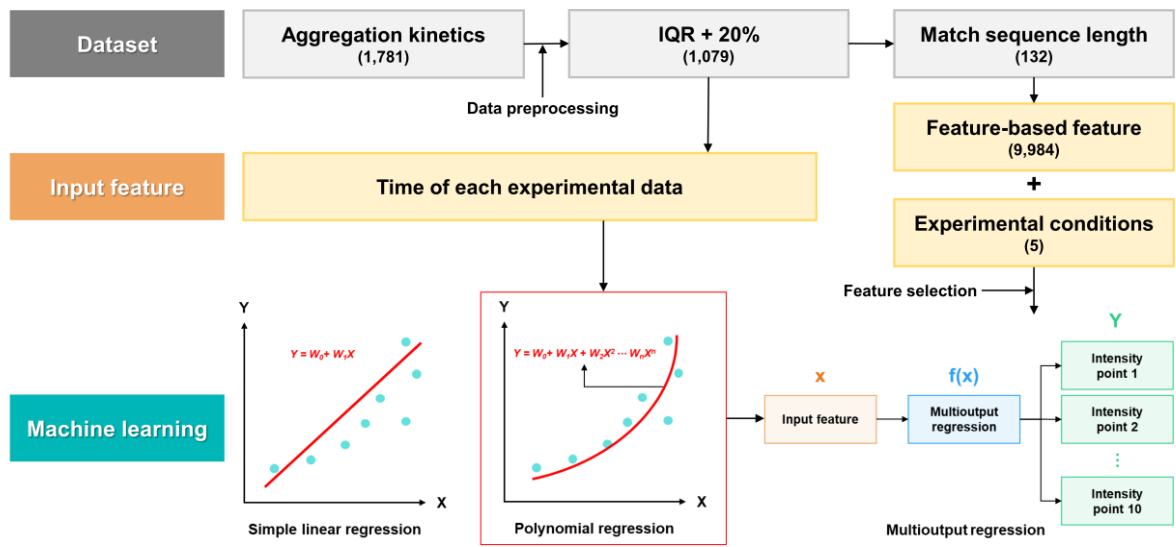

**Figure 4.** Flowchart of machine learning base on experimental CPAD 2.0 database

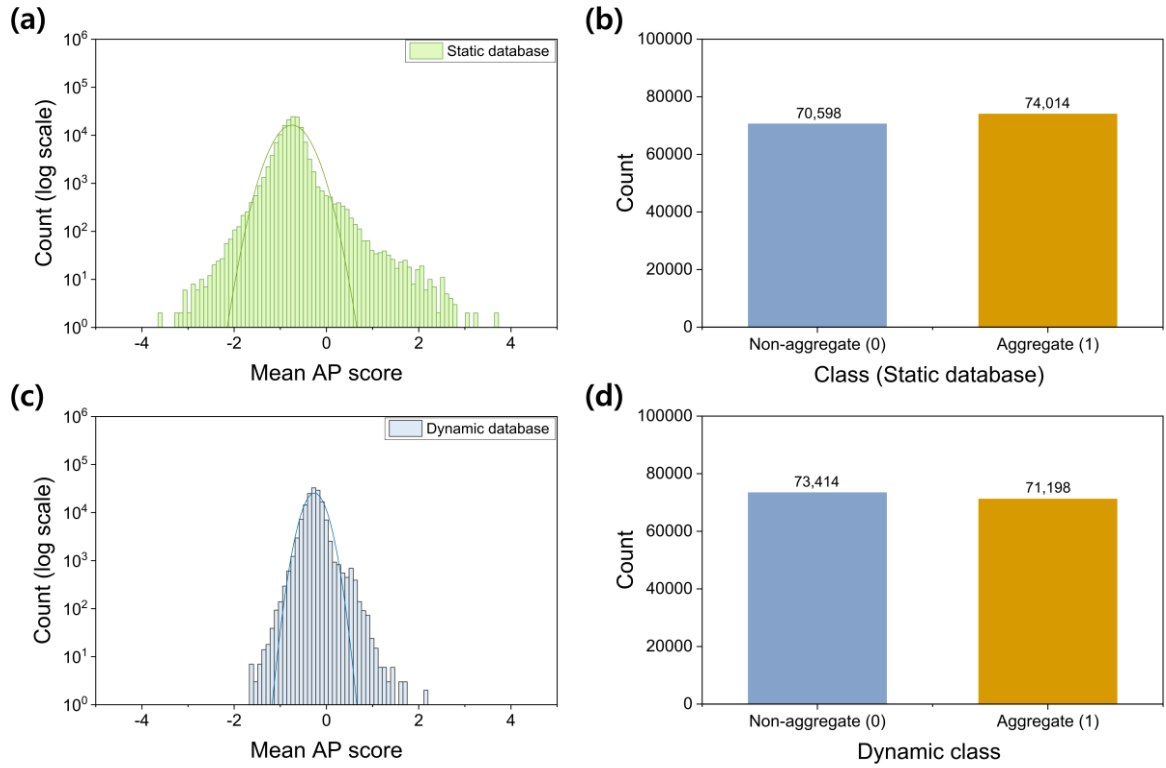

**Figure 5.** Distribution of static and dynamic database and class (Non-aggregate (0), Aggregate (1)) which divided by aggregation rate: **(a), (b)** static database, and **(c), (d)** dynamic database.

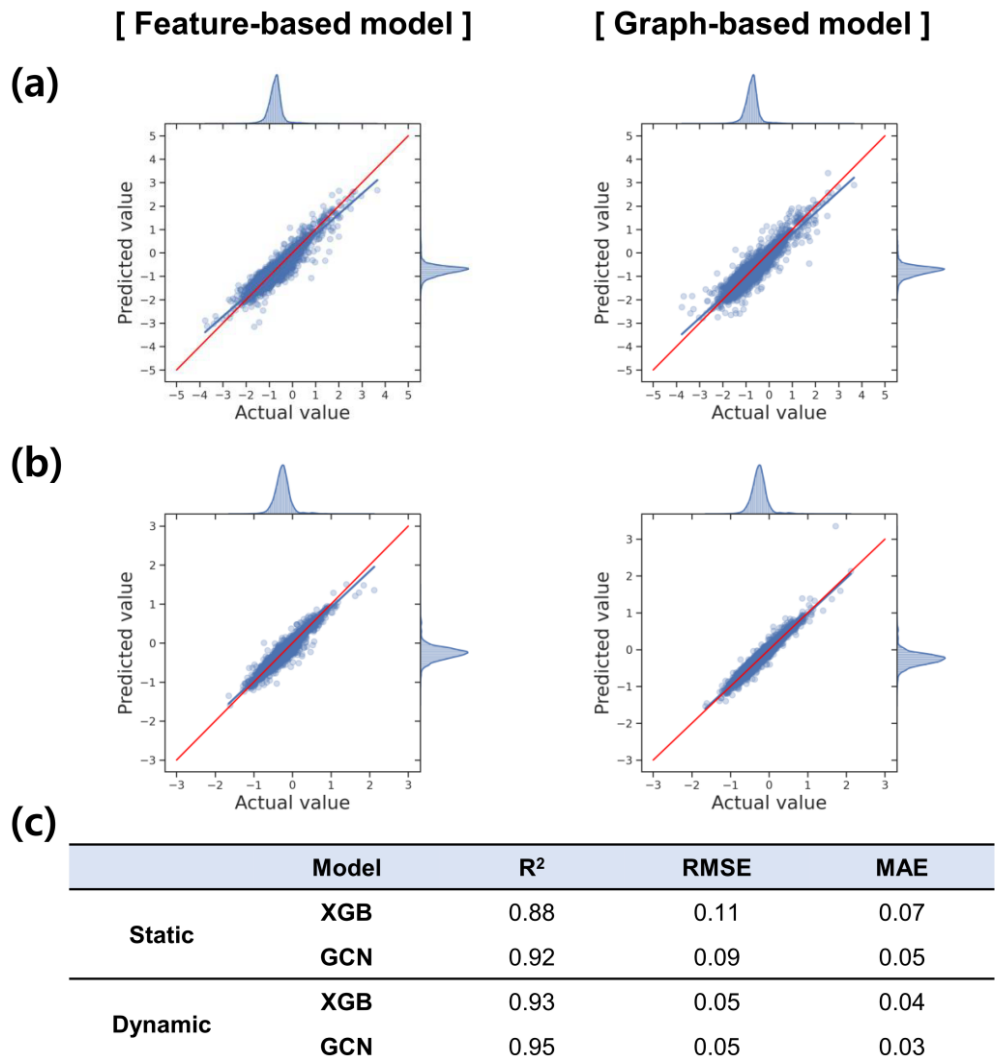

**Figure 6.** Scatter plot of regression result; **(a)** static database, **(b)** dynamic database. **(c)** Detail model performances of regression model.

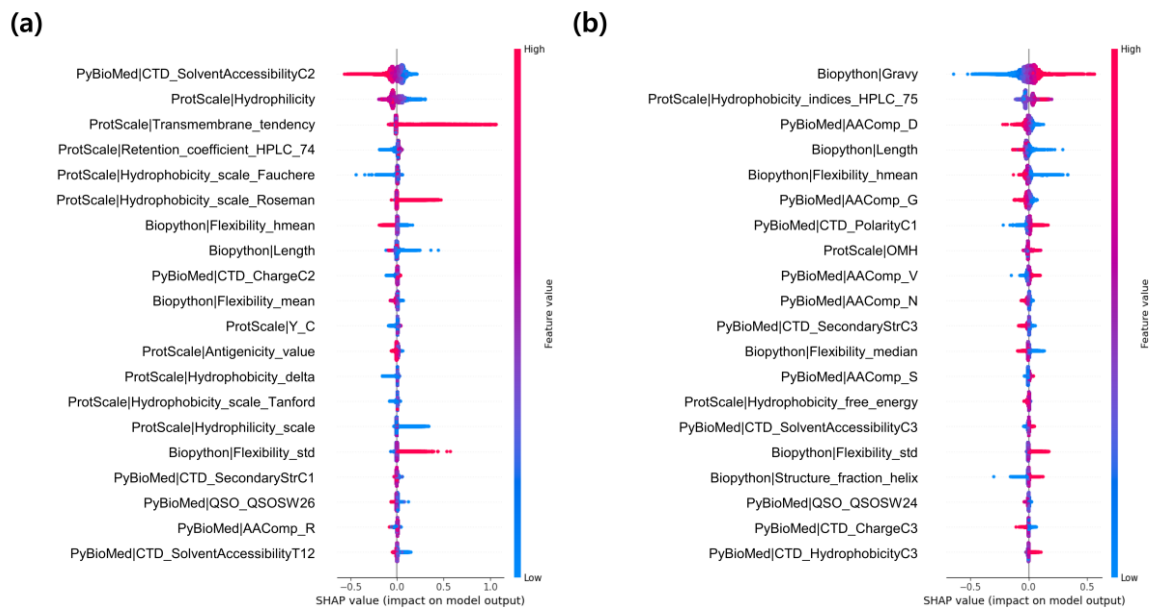

**Figure 7.** Regression SHAP value for **(a)** static database and **(b)** dynamic database.

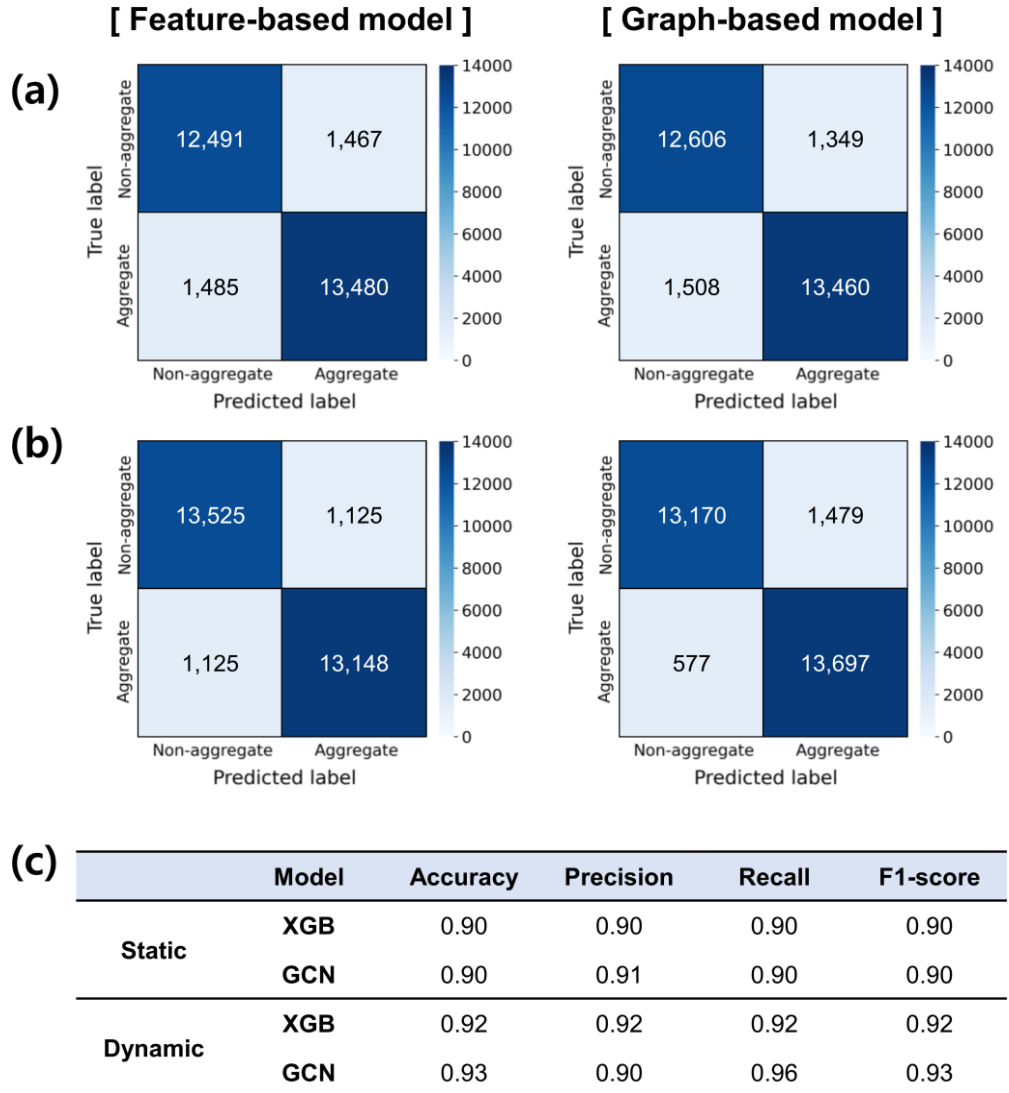

**Figure 8.** Confusion matrix of classification result; **(a)** static database, **(b)** dynamic database (Non-aggregate (Class : 0), Aggregate (Class : 1)). **(c)** Detail model performances of classification model.

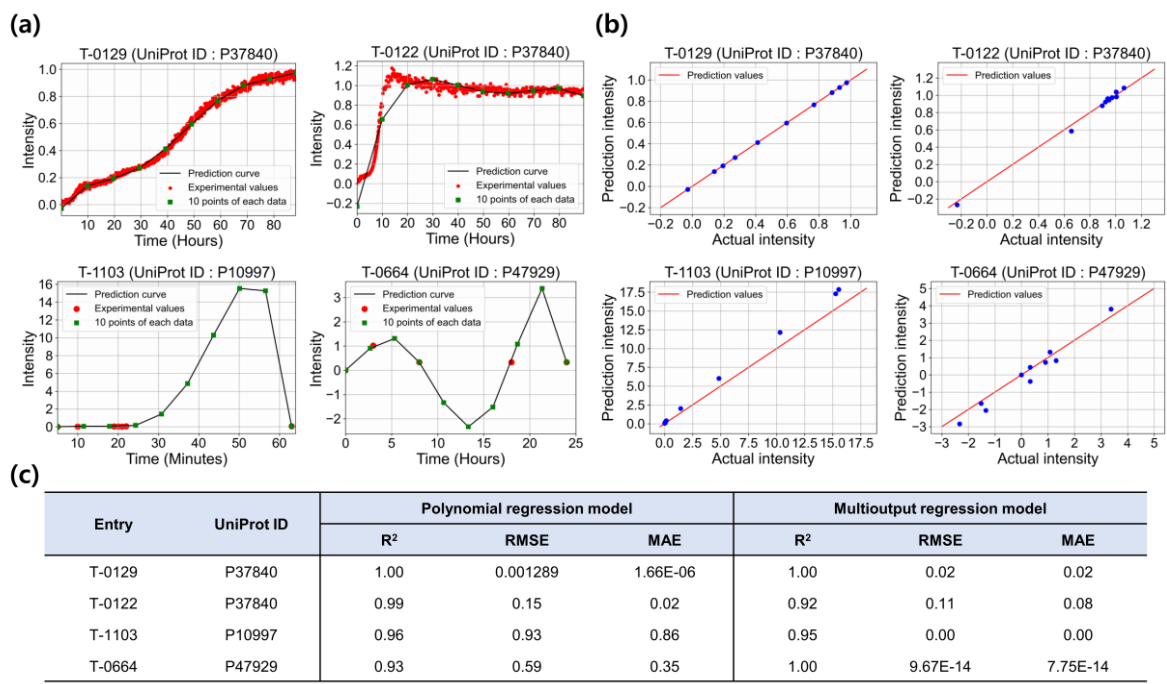

| Entry | UniProt ID | Polynomial regression model | | | Multioutput regression model | | |
|---|---|---|---|---|---|---|---|
| | | $R^2$ | RMSE | MAE | $R^2$ | RMSE | MAE |
| T-0129 | P37840 | 1.00 | 0.001289 | 1.66E-06 | 1.00 | 0.02 | 0.02 |
| T-0122 | P37840 | 0.99 | 0.15 | 0.02 | 0.92 | 0.11 | 0.08 |
| T-1103 | P10997 | 0.96 | 0.93 | 0.86 | 0.95 | 0.00 | 0.00 |
| T-0664 | P47929 | 0.93 | 0.59 | 0.35 | 1.00 | 9.67E-14 | 7.75E-14 |

**Figure 9.** Top four materials with the highest predictive performance in multioutput regression model. **(a)** The validation performance of the polynomial model and **(b)** multioutput regression model. **(c)** The performance of polynomial and multioutput regression model for top four proteins.